\documentclass[twocolumn,amsmath,amssymb,nofootinbib,superscriptaddress]{revtex4-2}
\usepackage{graphics,amssymb,amsmath,epsfig,color,textgreek}
\usepackage{graphicx}
\usepackage[dvipsnames]{xcolor}
\usepackage{dcolumn}
\usepackage{bm}
\usepackage[colorlinks=true,citecolor=cyan]{hyperref}
\hypersetup{colorlinks=true,citecolor=cyan,linkcolor=red,urlcolor=magenta}
\usepackage{braket}
\usepackage[normalem]{ulem}
\usepackage{cancel}
\usepackage{xfrac} 
\usepackage{amsfonts}
\usepackage{amsmath}
\usepackage{amssymb}
\usepackage{physics}
\usepackage{svg}
\usepackage{lipsum}

\newcommand{\trel}{t_{\mathrm{rel}}}
\newcommand{\tave}{t_{\mathrm{ave}}}
\renewcommand{\qq}{\mathbf{q}}
\newcommand{\kk}{\mathbf{k}}
\newcommand{\rr}{\mathbf{r}}
\newcommand{\pdag}{{\phantom\dagger}}
\newcommand{\tp}{t^\prime}

\newcommand{\LK}[1]{\textcolor{black}{#1}}

\begin{document}

\title{What do the two times in two-time correlation functions mean for interpreting tr-ARPES?
}

\author{J.~K.~Freericks}
\email{james.freericks@georgetown.edu}
\affiliation{Department of Physics, Georgetown University, 37th and O Sts. NW, Washington, DC 20057 USA}

\author{Alexander F. Kemper}
\email{akemper@ncsu.edu}
\affiliation{Department of Physics, North Carolina State University, Raleigh, North Carolina 27695, USA}

\date{\today}

\begin{abstract}
Time-resolved angle-resolved photoemission spectroscopy is one of the most powerful pump-probe measurements of materials driven far from equilibrium. Unlike the linear-response regime, where the frequency-dependent response function is independent of time, in a far-from-equilibrium experiment, the response function depends on two times in the time domain. In this work, we describe how one can use time-dependent frequency response functions and how they involve contributions from times that are near to each other. This implies that they should not be thought of as a frequency-dependent response at a single definite time. Instead, the Fourier uncertainty relations show that they involve contributions from ranges of times and must be interpreted in this light. We use this insight to help understand what time-resolved photoemission measurements actually measure.
\end{abstract}

\maketitle

\section{Introduction}

Much of experimental physics is based on measuring small perturbations of systems from the equilibrium state---a practice that is called linear-response. In this regime, we can measure frequency-dependent responses, such as reflectivity, photoemission spectroscopy and so on. Because systems in equilibrium are time translation invariant (or, if you prefer, homogeneous in time), one can directly measure frequency-dependent properties that do not depend on the time at which they are measured. 

Recently, there has been an increasing interest in nonequilibrium systems, which are far from equilibrium. These systems are prepared by providing a strong excitation (a pump pulse) to the system and measuring the response (a probe pulse) as a function of some time delay relative to the when the pulse was applied. While we might expect there to be similar frequency-dependent responses as functions of the delay time relative to the pump pulse, this concept is actually no longer well defined, because the system is no longer time-translation invariant (due to the pump). Just like one has an uncertainty relation between position and momentum, which can be thought of as arising from the Fourier transform that relates the two, frequency-dependent responses cannot be thought to occur at definite times in far from equilibrium systems---the Fourier transformation governs the question of what \textit{range} of times are involved in determining a frequency-dependent response function centered at some average time. The situation is often further complicated by the fact that many nonequilibrium response functions  depend on \textit{two times}, when a system is driven far from equilibrium. These concepts can be easily confused and misunderstood. 

In this paper, we will describe how to carefully analyze and understand what frequency measurements mean in the time domain (for systems driven into nonequilibrium). We focus our discussion on the problem of time-resolved photoemission. In this experiment, the system is excited by a low-photon-energy pump pulse and then a probe pulse (of a higher photon energy) is applied to photo-emit electrons. The electrons are then collected with both energy and momentum resolution (within a measurement time window). The theoretical goals for analyzing these experiments are to determine what this spectra looks like as a function of the delay time of the probe and to understand what these spectra are measuring.

\section{Two-time response functions}

When we make a measurement in physics, we typically measure a response
function.  That is, given some operator $\hat{B}(\rr^\prime,t^\prime)$ at position $\rr^\prime$ and time $t^\prime$, we find the joint expectation value
with another operator $\hat{A}(\rr,t)$ at another position $\rr$ and a later time $t$:
\begin{align}
    \chi^R(\rr,\rr^\prime;t,t^\prime) = -i\langle [\hat{A}(r,t), \hat{B}(r^\prime,t^\prime)]\rangle \theta(t-t^\prime).
    \label{eq:chi_neq}
\end{align}
The $\theta$ function (unit step function) ensures causality; that is, it ensures that $\hat{A}$ is measured after $\hat{B}$,
and the resulting correlation function is thus called a retarded correlation function (denoted by a superscript $R$). The commutator enters, because in quantum mechanics, both processes ($\hat{A}$ acting first on the state or $\hat{B}$ acting first) are allowed and we must include both contributions).
Common examples of these include the conductivity (a current-current correlation function
$\langle \hat{j}(\rr,t), \hat{j}(\rr^\prime,t^\prime)]\rangle$), spin-resolved neutron scattering response function (given by a spin-spin correlation
function $\sigma^{x,y,z}_{\alpha\beta} =\langle [\hat{S}_\alpha(\rr,t), \hat{S}_\beta(\rr^\prime,t^\prime)]\rangle$),
and angle-resolved photoemission. {\color{black} Note, that the generic term ``response function'' is often used for any correlation function defined in this fashion, even if not all such correlation functions can be easily measured by experiment.}

When the system is in equilibrium, it does not matter precisely when the experiment is performed; that is, we have time
translation invariance.  In this case, the correlation
function depends solely on the \emph{time difference} $t-t^\prime$ between the operators, also known as the relative time
$t_\mathrm{rel}$. Hence, 
\begin{align}
    \chi^R(\rr,\rr^\prime;\trel) = -i\langle [\hat{A}(\rr,\trel) , \hat{B}(\rr^\prime,0)] \rangle \theta(\trel).
    \label{eq:chi_eq_tr}
\end{align}
A similar consideration applies to spatial translation invariance in spatially homogeneous systems.

Measurements in equilibrium are commonly done in the frequency domain and in momentum space, which is obtained
from Eq.~\eqref{eq:chi_eq_tr} via Fourier transformation with respect to the relative time and the relative position, and satisfies
\begin{align}
    \chi^R(\qq,\omega) = -i\int^\infty_{-\infty} d\rr \int_0^{\infty} dt \langle [\hat{A}(\rr,t) , \hat{B}(0,0)]\rangle e^{i(\omega t- \qq \cdot \rr)}.
\end{align}

To illustrate how this works in detail, consider the specific example of angle-resolved photoemission spectroscopy (ARPES). 
ARPES involves making single-particle excitations out of the many-body
state, which are typically expressed using second quantized language; that is, using fermionic raising and
lowering operators $\hat{c}^\dagger$ and $\hat{c}$. The single-particle correlation functions are 
also termed Green's functions and denoted by $G$. ARPES specifically measures the
occupied states (denoted by $G^<$, more on this later), 
which we can express as
\begin{align}
    G^<_\sigma(\rr,\rr^\prime;t,t^\prime) &= i  \langle \hat{c}^\dagger_\sigma(\rr^\prime,t^\prime) \hat{c}^\pdag_\sigma(\rr,t)\rangle\\
    &=i\mathcal{Z}^{-1}\text{Tr}\left \{e^{-\beta \hat{\mathcal H}(t_0)}\hat{c}^\dagger_\sigma(\rr^\prime,t^\prime) \hat{c}^\pdag_\sigma(\rr,t)\right \};\nonumber
\end{align}
that is, the probability of a single-particle excitation making its way from $(\rr^\prime,t^\prime)$ to $(\rr,t)$. Here, the partition function is defined to be $\mathcal{Z}={\rm Tr} \left \{e^{-\beta \hat{\mathcal H}(t_0)}\right \}$ and the angle brackets are defined by the second line; the time $t_0$ is a reference time when the system is initially in equilibrium, or equivalently it is a time \textit{before} the nonequilibrium perturbation is turned on. The operators are expressed in the Heisenberg representation. Note that the lesser Green's function has no causal structure, nor does it have time-ordering or commutators in its definition; in addition, we are ignoring so-called matrix-element effects, which can play important roles, but are not discussed further in this work. Using
spatial translation invariance, we transition to Bloch states with quasi-momentum $\kk$, so that the lesser Green's function is given by
\begin{align}
    G^<_\sigma(\kk;t,t^\prime) = i \langle \hat{c}_{\kk\sigma}^\dagger(t^\prime) \hat{c}_{\kk\sigma}^\pdag(t)\rangle.
    \label{eq:gless_eq}
\end{align}
{\color{black}In this work, we restrict to a single-band model for the electrons.}

To gain some insight into this correlation function, we write out the explicit expectation value in Eq.~\eqref{eq:gless_eq},
\begin{align}
    G_{\kk\sigma}^<(t,t^\prime) &= 
    i    \sum_\gamma \rho_\gamma \langle \Psi_\gamma |  \hat{c}_{\kk\sigma}^\dagger(t^\prime) \hat{c}_{\kk\sigma}^\pdag(t) | \Psi_\gamma\rangle \\
    &= 
    i    \sum_\gamma \rho_\gamma \langle \Psi_\gamma |  \hat{U}(t_0,t^\prime) \hat{c}_\kk^\dagger \hat{U}(t^\prime,t) \hat{c}_\kk^\pdag \hat{U}(t,t_0) | \Psi_\gamma\rangle \nonumber
\end{align}
where $\rho_\gamma=\exp(-\beta E_\gamma)/\mathcal{Z}$ is the thermal weight of the energy eigenstate $|\Psi_\gamma\rangle$ (which satisfies $\hat{\mathcal{H}}(t_0)|\Psi_\gamma\rangle=E_\gamma|\Psi_\gamma\rangle$). In the second line, we have introduced the
time evolution operator $\hat{\mathcal H}$, which satisfies $i\partial_t\hat{U}(t,t_0)=\hat{\mathcal{H}}(t)\hat{U}(t,t_0)$; we also used the facts that this is a unitary operator and that $\hat{U}(t',t_0)=\hat{U}(t',t)\hat{U}(t,t_0)$.

The expectation value above can be viewed as the inner product between two states at time $t^\prime$,
\begin{align}
    |\Phi_1\rangle &= \hat{c}_{\kk\sigma} \hat{U}(t^\prime,t_0) |\Psi_\gamma\rangle \\
    |\Phi_2\rangle &= \hat{U}(t^\prime,t) \hat{c}_{\kk\sigma} \hat{U}(t,t_0) |\Psi_\gamma\rangle.
\end{align}
That is, to obtain $|\Phi_1\rangle$ we propagate $|\Psi_\gamma\rangle$ to time $t^\prime$ and remove a particle
with momentum $\kk$ and spin $\sigma$; to obtain $|\Phi_2\rangle$ we propagate $|\Psi_\gamma\rangle$ to time $t$, remove a particle (also of momentum $\kk$  and spin $\sigma$),
and further propagate it from $t$ to $t^\prime$; if $t>t'$ the time evolution would be backwards in time.  Notice that if single-particle excitations are
not an eigenstate of the many-body system, the action of $\hat{U}(\tp,t)$ will cause the excitation to
spread in the Hilbert space. The Green's function is then formed by a weighted average of all of these overlaps between the two modified states. The final result has an amplitude and a complex phase representing the measurement.

\subsection{Equilibrium}
When there is no explicit time dependence in the Hamiltonian, that is we are in equilibrium, the inner product becomes
\begin{align}
  \langle \Psi_\gamma |  \hat{c}_{\kk\sigma}^\dagger(t^\prime) \hat{c}_{\kk\sigma}^\pdag(t) | \Psi_\gamma\rangle
  =& \langle \Psi_\gamma | e^{-i\hat{\mathcal H}\times(t_0-\tp)} \hat{c}_{\kk\sigma}^\dagger \nonumber \\
  &e^{-i\hat{\mathcal H}\times(\tp-t)} \hat{c}_{\kk\sigma}^\pdag  e^{-i\hat{\mathcal H}\times(t-t_0)} | \Psi_\gamma\rangle
\end{align}
where $\hat{\mathcal H}=\hat{\mathcal H}(t_0)$ is the time-independent Hamiltonian; the terms in the exponent in parenthesis are \textit{multiplying} the Hamiltonian operator, not arguments of the Hamiltonian representing its time dependence. We can further insert a complete set of
states (indexed by $\lambda$) to the right of the creation operator to find that
\begin{align}
    G_{\kk\sigma}^<(t-\tp) =& i \sum_{\gamma\lambda} \rho_\gamma |\langle \Psi^N_\gamma | \hat{c}_{\kk\sigma}^\dagger | \Psi^{N-1}_\lambda \rangle |^2 
    \nonumber \\ 
    &\times e^{-i(E_\gamma^N - E_\lambda^{N-1})(t-\tp)};
\end{align}
{\color{black}this expression is usually called the Lehmann representation~\cite{bruus2004many}.}
Here, the $E_\gamma^N$ denote the eigen-energies for the state $|\Psi_\gamma\rangle$ in the $N$-particle sector, and
similar for $E_\lambda^{N-1}$ in the $N-1$-particle sector. Note that this Green's function determines an electron removal spectrum.

\subsubsection{Non-interacting systems}
\begin{figure}
    \centering
    \includegraphics[width=\columnwidth]{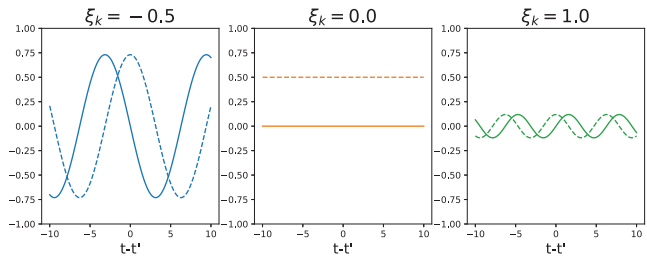}
    \caption{Lesser Green's function in equilibrium
    for non-interacting electrons as a function
    of relative time $\trel$. Here, the temperature satisfies $T=1$.}
    \label{fig:Gless_eq_time}
\end{figure}

In the limit where removing a single particle does not affect the eigen-energies of the remaining particles,
this further simplifies to
\begin{align}
    G_{\kk\sigma}^<(t-\tp) &= i f(\xi_\kk) e^{-i\xi_\kk(t-\tp)},
\end{align}
where $f(\xi_\kk)=1/(1+e^{\beta\xi_\kk})$ is the Fermi-Dirac distribution function, and $\xi_\kk$ is the quasiparticle energy. Note that, as expected,
this is a function of $\trel=t-\tp$ only; moreover, the Fermi function appears, indicating that
we are measuring the occupied states.  Fig.~\ref{fig:Gless_eq_time} shows the dependence of $G^<_{\kk\sigma}(t,\tp)$ as
a function of $\trel$ for a few quasiparticle energies. 
Since it has no absolute time
dependence, it is a constant function along the \emph{average} time $\tave = \frac{1}{2}(t+\tp)$ axis.

\begin{figure}
    \centering
    \includegraphics[width=0.45\columnwidth]{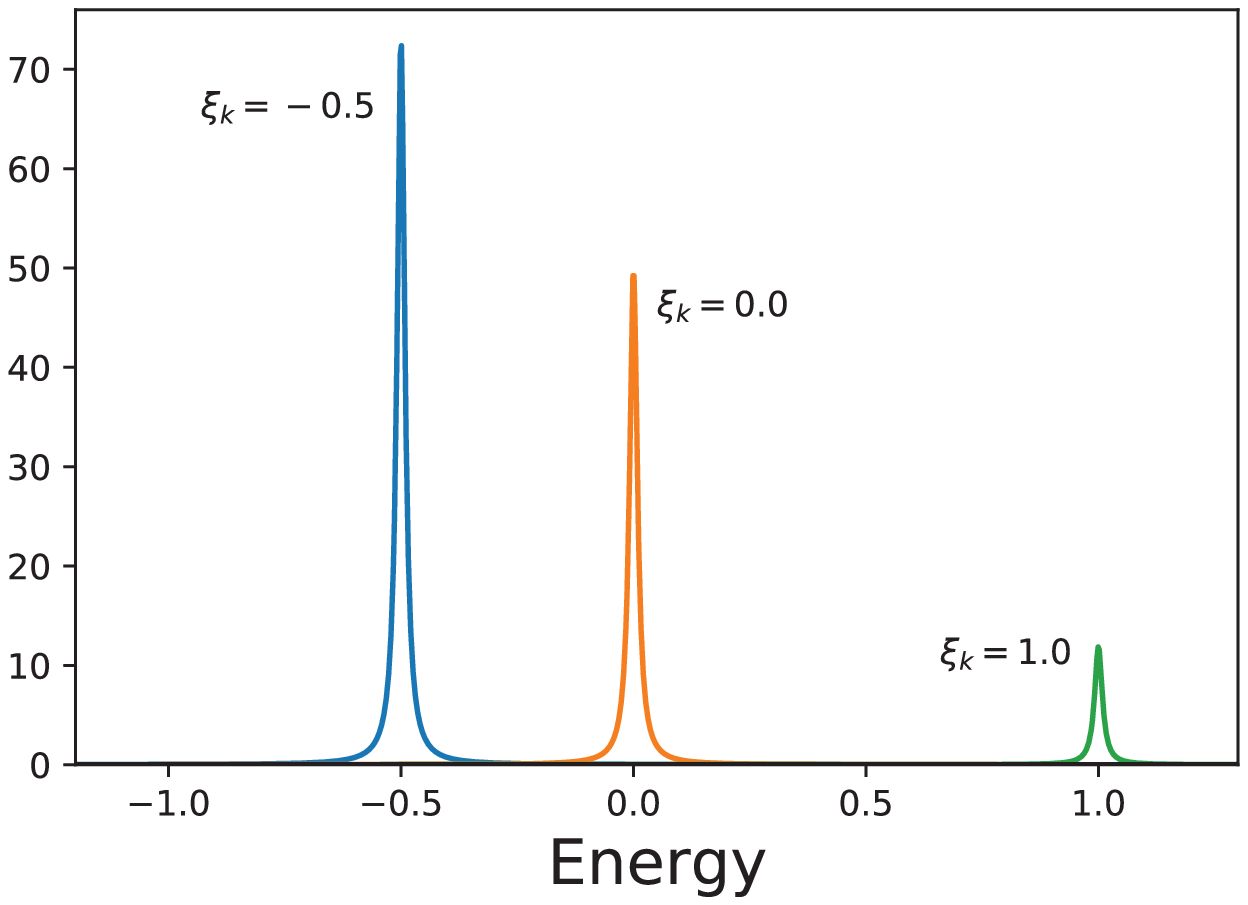}
    \includegraphics[width=0.45\columnwidth]{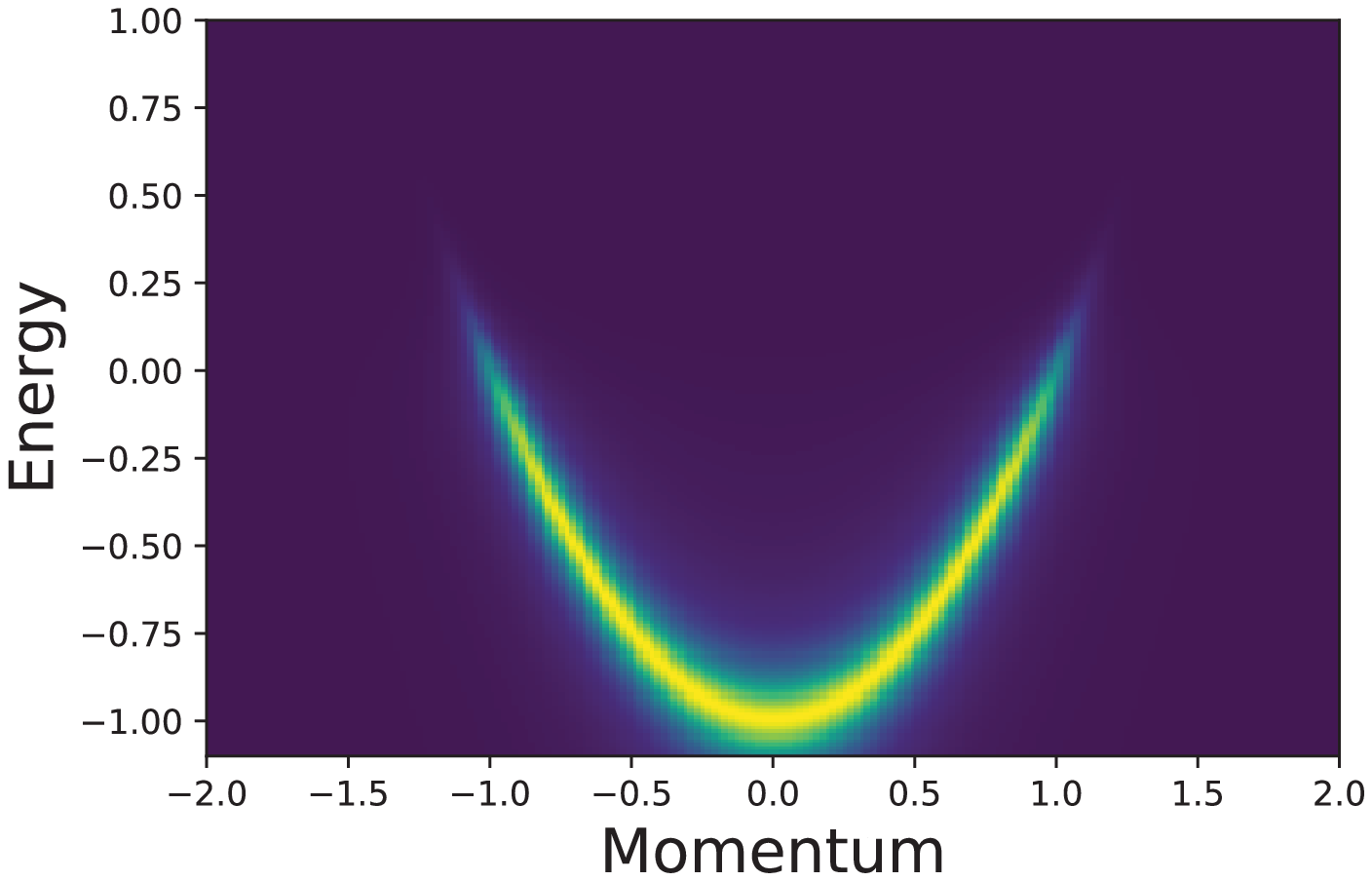}
        \caption{Left: Lesser Green's function in equilibrium
    for non-interacting electrons as a function
    of frequency $\omega$. Right: False color
    plot of the equivalent band structure. Here, $T=1$. Both panels include an artificial broadening $\Gamma$ of 0.01 in
    order to be able to represent the otherwise infinitely sharp spectral lines.
    }
    \label{fig:Gless_eq_freq}
\end{figure}

In the frequency (energy) domain, a simple Fourier transform along $\trel$ yields
\begin{align}
    G^<_\kk(\omega) &= 2\pi i f(\xi_\kk) \delta(\omega - \xi_\kk) \\
    &= 2\pi i f(\omega) \delta(\omega - \xi_\kk)
\end{align}
where $\delta(x)$ is the Dirac delta function. Thus, the lesser Green's function (and ARPES)
is peaked right at the quasiparticle spectrum, or at the band energies (see Fig.~\ref{fig:Gless_eq_freq}).

We could have equally well considered the retarded (or ``causal'' version) of the Green's function, which yields (in the same equilibrium  and non-interacting limit)
\begin{align}
    G^R_{\kk\sigma}(t-\tp) &= -i\theta(\trel) e^{-i\xi_\kk(t-\tp)}, \\
    G^R_{\kk\sigma}(\omega) &= \frac{1}{\omega - \xi_\kk + i0^+},
\end{align}
that is, the well-known expression for the noninteracting equilibrium retarded Green's function.  From here, we can also read off the relationship
between the retarded and lesser components (in equilibrium),
\begin{align}
    G^<_{\kk\sigma}(\omega) = -2i f(\omega) \mathrm{Im} G^R_{\kk\sigma}(\omega)
    \label{eq:fluc_diss} 
\end{align}
also known as the \emph{fluctuation-dissipation theorem.} While this
has been explicitly demonstrated in the non-interacting case, this form holds 
in general for interacting systems in equilibrium.

\subsubsection{Interacting systems}

\begin{figure}
    \centering
    \includegraphics[width=0.9\columnwidth]{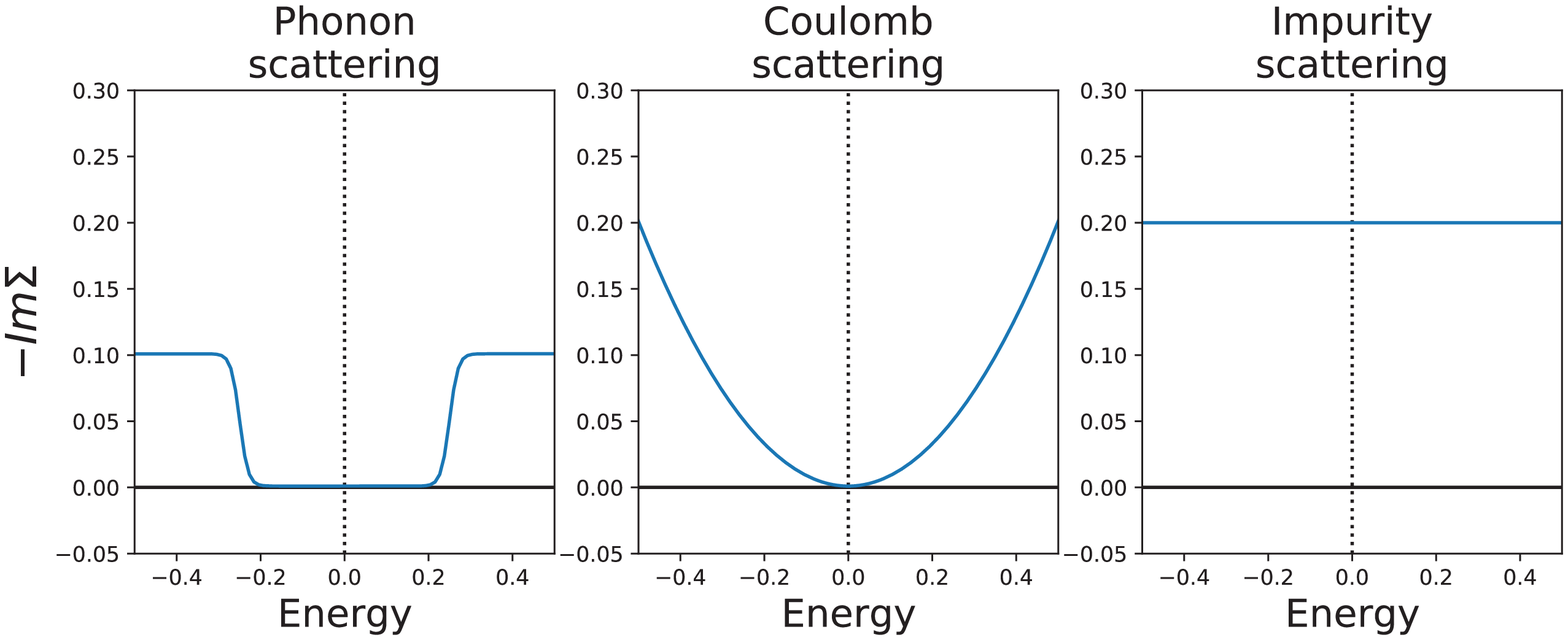}
    \includegraphics[clip=true,trim=0 0 0 45,
    width=0.9\columnwidth]{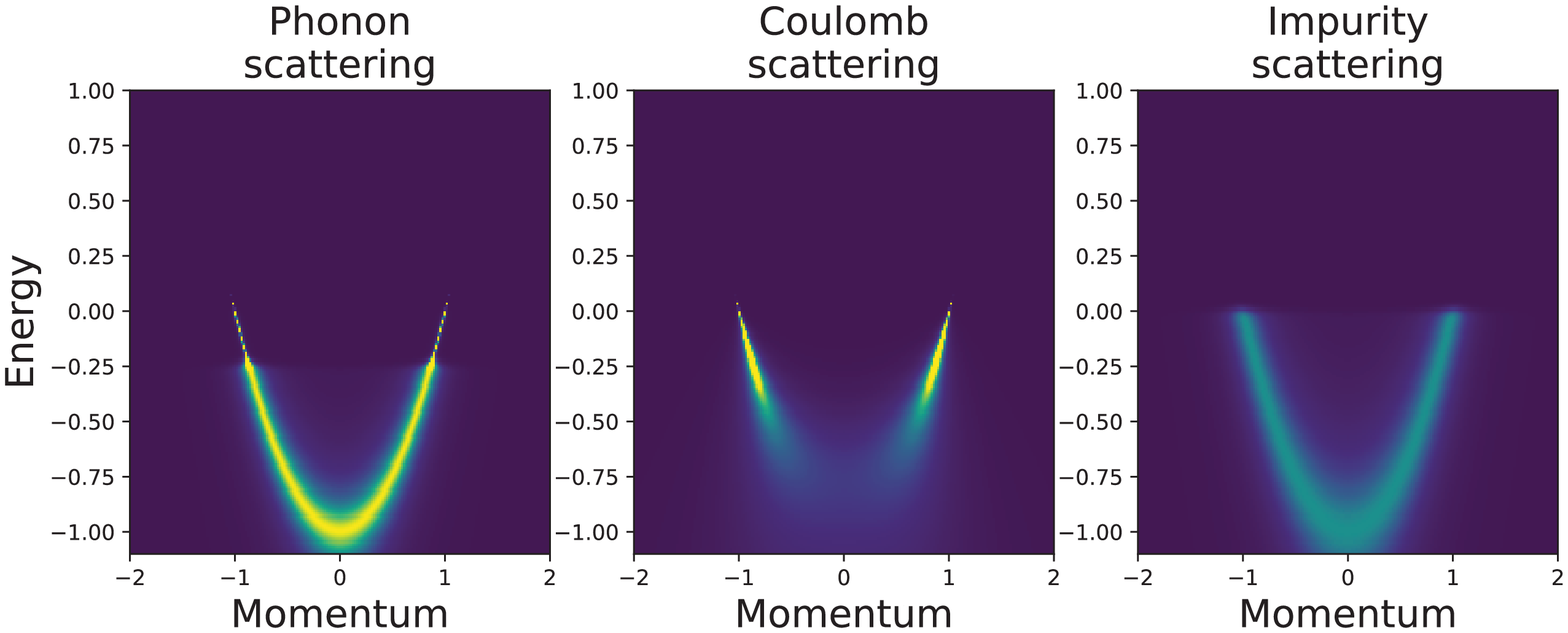}
        \caption{Top: Local (momentum-summed) self-energies for various scattering processes. Bottom:
        resulting equilibrium ARPES spectra.}
    \label{fig:sigmas}
\end{figure}

When electrons are interacting, single-particle excitations (as found in $\ket{\Phi_{1}}$ and $\ket{\Phi_{2}}$)
are no longer eigenstates, and thus $\ket{\Phi_2}$ will
spread out in Hilbert space as time goes on. We may expect the overlap 
$\braket{\Phi_1}{\Phi_2}$ will decay as $|t-\tp|\rightarrow\infty$. If the only affect is the decay over relative time, this changes the lesser Green's function to
\begin{align}
    G_{\kk\sigma}^<(t-\tp) &= i f(\xi_\kk) e^{-i\xi_\kk(t-\tp)} e^{-\frac{1}{2}\Gamma|t-\tp|}.
\end{align}
In the frequency domain, this modifies the lesser Green's function to the form
\begin{align}
    G^<_{\kk\sigma}(\omega) = -2 i f(\omega) \mathrm{Im } \frac{1}{\omega - \xi_\kk - i\Gamma}.
\end{align}
Here, $\Gamma$ plays the role of the \emph{line width}.
In more general cases, $\Gamma$ gets replaced by the \emph{self energy} $\Sigma_\kk(\omega)$,
which encodes the effects of the interactions on the single-particle excitations (corresponding to a frequency-dependent change in the energy of the excitation via its real part and a frequency-dependent change to the linewidth via its imaginary part).
Since the self-energy encodes information about the interactions, the self-energy typically exhibits features that
reflect specific features of different forms of electron interactions. For example, Fig.~\ref{fig:sigmas} shows the typical self-energy and resulting
spectra due to electron-phonon (el-ph) coupling, Coulomb (el-el) interactions, and impurity scattering (el-imp)
in the approximation where the self-energies are local.
{\color{black}We use Migdal-Eliashberg theory for the electron-phonon self-energy (in the Holstein  model), the first-Born approximation for the impurity scattering, and a second-order perturbation theory for the Coulomb interaction (in the Hubbard model).}

When
multiple interactions are present, the various components typically add according
to Matthiessen's rule, 
\begin{align}
    \Sigma = \Sigma_\mathrm{el-ph} + \Sigma_\mathrm{el-el} + \Sigma_\mathrm{el-imp}.
\end{align}
Separating
the self-energy into these different components has led to significant insight into the physics
of strongly correlated matter (see e.g.~\cite{choi2018pin}).
Going back to the time domain, the correlation function decay rate
is related to the self-energy via
\begin{align}
    \Gamma = -2\mathrm{Im }\Sigma
    \label{eq:tau_sigma}
\end{align}
when there is a single decay rate for all momenta.
This relation becomes more complex when the self-energy has energy and momentum dependence (which is what usually occurs).

\subsection{Non-equilibrium}
\begin{figure}
    \centering
    \includegraphics[width=0.9\columnwidth]{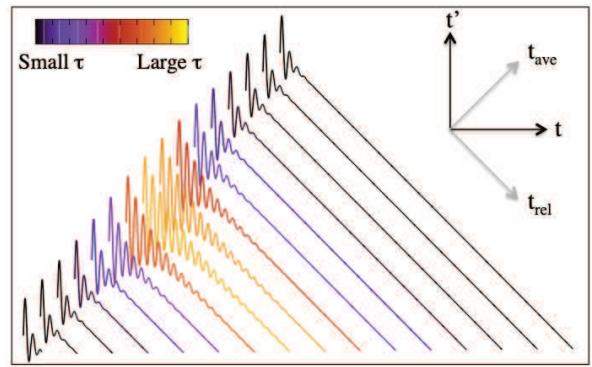}
        \caption{Illustration of the two-time retarded Green's function
        $G^R(t,t')$. Due the pump, the decay rate $\tau^{-1}$
        undergoes a decrease and subsequent increase as a function of $\tave$, which is reflected in the longer
        range of the oscillations. The figure also illustrates
        the relationship between the $t,t^\prime$ and
        $\tave,\trel$ directions.}
    \label{fig:gless_rot}
\end{figure}

When studying correlation functions out of equilibrium (say, for pump-probe experiments), time-translation invariance
is broken by the presence of a pump pulse, which sets an absolute reference point in time.
This implies that we can no longer work with $\trel$ only, and we have to account for the full
dependence on both $t$ and $\tp$ as illustrated in Eq.~\ref{eq:chi_neq}. We can, however,
rotate the time axes to the diagonals, which correspond to \emph{relative} and \emph{average} times $\trel$ and $\tave$ and provide additional
insight to the dynamics (see Fig.~\ref{fig:gless_rot}). Now, in equilibrium, there is no change in the spectra
along the $\tave$ axis; this steady state is achieved through a balance of scattering
rates for each state. Out of equilibrium, there are dynamics along $\tave$,
however they are expected to be different from those occurring along the $\trel$ axis; in general, the dynamics along the two directions do not separate.

One situation where we can make some progress is when
\begin{enumerate}
    \item We are making measurements long after the pump, \textit{i.~e.~} when the external driving field has returned to zero.
    \item The time dependence along the measurement time ($\tave$) is slow compared to the time dependence along $\trel$.
\end{enumerate}
In this case, the time dependence along $\tave$ may be mapped onto the parameters of the
correlation function. For example, the scattering rate $\Gamma$ may acquire an average
time dependence: $\Gamma(\tave)$:
\begin{align}
    G^<_{\kk\sigma}(t,\tp) = i f(\xi_\kk) e^{-i\xi_\kk \trel} e^{-\Gamma(\tave)|\trel|},
\end{align}
as illustrated in Fig.~\ref{fig:gless_rot}. We can subsequently take the Fourier transform along $\trel$
(which is allowed because the support of the signal is limited to a regime where there is essentially 
no $\tave$ dependence) approximating the average time dependence of the parameters to be fixed as $\trel$ varies. Typical parameters that are modeled to change as a function
of $\tave$ are the electronic/lattice temperature, scattering rates (in a Drude formalism),
order parameters (\textit{e.~g.~}superconducting gaps), or lattice vibration (phonon) frequencies.
In a few cases, the self-energy itself has been treated as weakly average-time dependent, although the times scales
are harder to separate in the self-energy.

The positive side of using (average) time dependent parameters is that it makes a direct
connection to equilibrium physics and is easy to interpret. However, it also means that
we are forcing nonequilibrium behavior into an effective equilibrium form, and hence it must be limited. Moreover, it is difficult
to tell when these approximations are valid, and when they break down, but it is also clear that one would not expect that the average time dependence of a parameter would always be fixed as the relative time varies. So this approximation must become inaccurate at some point. \LK{In fact, it was shown both experimentally~\cite{yang2015inequivalence} and theoretically~\cite{lex_prx} that a direct connection between the out-of-equilibrium average time
decay rate $\Gamma(\tave)$ and the equilibrium self-enery $\Sigma$ breaks down except in very simple cases (see e.g.
Refs.~\cite{allen1987theory,kemper2013mapping,sentef2013examining,sangalli2015ultra}).
We discuss this further in Sec.~\ref{sec:pop}.}

\section{Gentle introduction to Keldysh Green's Functions}

In Sec.~II, we have introduced the ``lesser''
and ``retarded'' components of the Green's function.
These different components arise from the various orderings
of the creation/annihilation operators in the Green's 
function. For example, the greater/lesser components
are given by
\begin{align}
    G_{\kk\sigma}^>(t,\tp) &= -i\langle \hat{c}^\pdag_{\kk\sigma}(t) \hat{c}^\dagger_{\kk\sigma}(\tp) \rangle \\
    G_{\kk\sigma}^<(t,\tp) &= i\langle  \hat{c}^\dagger_{\kk\sigma}(\tp)\hat{c}^\pdag_{\kk\sigma}(t) \rangle.
\end{align}
These correspond to the unoccupied/occupied states,
respectively, which can be seen from the fixed ordering
of the operators. Alternatively, these are single-particle
hole/electron excitations. The 
names ``lesser'' and ``greater'' come from the ordering of
the times along the Keldysh contour described below. Note that these Green's functions are not causal or acausal, but they exist for all values of the two times.
The other relevant component, the retarded Green's function,
combines the two to form the full spectrum (including the causal  unit step function)
\begin{align}
    G_{\kk\sigma}^R(t,\tp) &= \theta(t-\tp) \big ( G_{\kk\sigma}^>(t,\tp) -   G_{\kk\sigma}^<(t,\tp)\big )\\
    &=-i\theta(t-\tp)\langle \{\hat{c}^\pdag_{\kk\sigma}(t), \hat{c}^\dagger_{\kk\sigma}(\tp)\} \rangle.
\end{align}
In the frequency
domain, the spectra of the greater/lesser Green's functions
correspond to the unoccupied/occupied spectra, and the spectrum
of the retarded Green's function to the full spectrum. In other words, the retarded/advanced Green's functions tell us about the density of quantum states, while the lesser/greater Green's functions tell us how those quantum states are occupied. This means the retarded/advanced Green's functions usually relax to their steady-state values before the lesser/greater Green's functions do so.

Since photoelectron spectroscopy measures the occupied spectrum
(both in and out of equilibrium), the relevant Green's function
is always the lesser one. However, it is sometimes useful
think about changes in the spectrum and the occupations separately. 
\LK{This does not always work; the 
the spectrum can change as
the occupations change in a many-body interacting system~\cite{tuovinen2020comparing},
or to put this another way, there may be significant self-energy effects on the spectrum. However,}
when it does work a useful framework
to treat this is the generalized Kadanoff-Baym approximation
(GKBA). In this case, we can write $G_{\kk\sigma}^<$ schematically via
\begin{align}
    G_{\kk\sigma}^<(t,t') \approx -G_{\kk\sigma}^R(t,t') \rho_{\kk\sigma}^<(t') + G_{\kk\sigma}^A(t,t') \rho_{\kk\sigma}^<(t)
\end{align}
with $\rho_{\kk\sigma}^<(t) = -iG_{\kk\sigma}^<(t,t)$ is an effective time-dependent
density for momentum $\kk$ and spin $\sigma$. This is useful for \textcolor{black}{some} calculations; \textcolor{black}{but has a mean-field-like character to its spectral moments as described in Ref.~\cite{freericks_moments}}. It \textcolor{black}{corresponds to} some
additional approximations, similar to the quantum Boltzmann equation, which we will discuss below.



\section{tr-ARPES from $G^<$}
Going beyond an approximate equilibrium based on slowly varying time-dependent
parameters requires obtaining the full, two-time dependent
$G^<_{\kk\sigma}(t,\tp)$; this is usually accomplished via numerical
means as closed form expressions are difficult to obtain for interacting systems.
The numerical approaches include exact diagonalization
(appropriate for small clusters), non-equilibrium Dyson
equation solvers~\cite{kemper2014effect}, embedding methods such as nonequilibrium
dynamical mean field theory~\cite{freericks2006nonequilibrium,aoki2014nonequilibrium}, or time domain density matrix
renormalization group~\cite{Feiguin2013b}. Each method has their own pros
and cons, but eventually they all evaluate Eq.~(\ref{eq:gless_eq})
in one way or other. After this is done, time-resolved ARPES
is obtained via post-processing. Following Freericks \textit{et al.}~\cite{freericks2009theoretical},
the formal expression for the time-resolved ARPES signal
is found to be a Gaussian-windowed Fourier transform along $\trel$,
with the window $s(t-t_0)$ centered around $\tave=t_0$:
\begin{align}
    I(\kk,\omega,t_0) = 
    \int_{-\infty}^\infty dt
    d\tp
    G_\kk^<(t,\tp) s(t-t_0) s(\tp-t_0) e^{-i\omega\trel},
\end{align}
where $s(t) = \exp\left(-t^2/2\sigma^2\right)$ and $\sigma$ is the width
of the Gaussian window that sets the trade-off between energy and time
resolution (one sums over both spin components as well, not shown, when the spin of the electron is not measured). A small window provides good time resolution, but is
only able to resolve the fastest (highest energy) oscillations
and is thus limited in energy resolution. Conversely, a wide
window will smear the dynamics along $\tave$, but is able
to resolve small energy features. 
When multiple bands are present, that is, the Green's
function has band (or orbital) indices $G^<_{\kk,a,b,\sigma}$,
band (orbital) dependent matrix elements may play a role. Note that the probe envelope functions depend on $t$ and $\tp$, respectively, which mixes the average and relative time dependence. This governs why one cannot think of spectra as occurring at a specific time.


Explicit calculations are always performed in a specific gauge. While the total photoemission signal {\color{black}calculated in this gauge} is gauge-invariant, the ARPES {\color{black} calculation} is not, and hence one needs to pay attention to gauge-invariance properties of the photoemission spectra. For single-band tight-binding-based models, the solution has been to {\color{black}instead} use the so-called gauge-invariant Green's function to calculate the response function~\cite{gi_gf}. When one works with multiple bands, the situation is much more complicated, as matrix element effects must be explicitly taken into account. Recent work has shown that this can be handled, at least in principle, but it can greatly complicate the formalism~\cite{sentef_gi}. {\color{black}Of course, even for single-band models, the gauge-invariance properties become more complicated when one includes matrix-element effects in the expressions used to determine the theoretical signal.}

\section{Observed dynamics}

With the conceptual basis of time-resolved ARPES in hand,
we now turn to the observed dynamics. In this section, we illustrate some of the phenomena that can be described by model-system calculations using advanced numerical methods. 

\subsection{Population dynamics}
\label{sec:pop}
The primary
observable in trARPES tends to be the population dynamics;
experiments are good at tracking the number of carriers in the
(otherwise) unoccupied states. Beyond the initial
identification of the unoccupied band structure (made visible by non-equilibrium pumping into thermally unoccupied states), the second
quantity of interest is the transfer of the electron population from one
state to another. For example, experiments on a topological
insulator Bi$_2$Se$_3$ showed that while electrons rapidly disappear
from the bulk conduction band, the dynamics of the topological
surface state are much slower~\cite{sobota2012ultrafast}. More detailed analysis
may reveal intricate details of the population dynamics, which
has been applied to a wide variety of systems 
\cite{cortes2011momentum,smallwood2012tracking,yang2015inequivalence,na2019direct,gierz2013snapshots,tanimura2019dynamics}

A question arose from this work: what do the obtained dynamics (typically
decaying exponentials) correspond to?  This particular point
is where the connection to equilibrium breaks down; the
decay rate along $\trel$, which is related to the equilibrium
self-energy via Eq.~(\ref{eq:tau_sigma}), is not the same as the 
population decay rate(s) (long $\tave$) observed with trARPES. This divergence
was pointed out experimentally by Yang \textit{et al.}~\cite{yang2015inequivalence}., who
demonstrated that in the high-$T_c$ cuprates the (equilibrium)
line width was much larger than the population decay rates.

We can gain some insight into this question by considering
well-defined quasi-particles (which exist when
the line width is not large) some time after the pump.
First, let us assume the 
relationship between the
occupations and the spectrum (the fluctuation-dissipation
theorem) above in Eq.~(\ref{eq:fluc_diss}) holds, but with
a $\tave$ dependence as
\begin{align}
    G^<_{\kk\sigma}(\tave,\omega) = -2i f^G_\kk(\tave) \mathrm{Im} G^R_{\kk\sigma}(\tave,\omega)
\end{align}
where $f^G_\kk(\tave)$ is the occupation of the state
$\kk$ at time $\tave$; it is a generalization of the
Fermi function to an arbitrary nonequilibrium occupation function. We write a similar relation for the self-energy, with $f_\kk^\Sigma(\tave)$.
For on-shell quasi-particles (i.e. at $\omega=\xi_\kk$),
the population dynamics then obey the differential equation~\cite{lex_jkf_opt}
\begin{align}
    \frac{dn_\kk(\tave)}{d\tave} =& 4 
    \mathrm{Im}\left[\Sigma^R(\tave,\xi_\kk) \right]
    \mathrm{Im}\left[G^R(\tave,\xi_\kk) \right] \nonumber \\
    &\times \left[ f^G(\tave,\xi_\kk) - f^\Sigma(\tave,\xi_\kk) \right].
\end{align}
The quantities in this equation are $n_\kk(\tave)$ ---
the population in momentum state $\kk$, the spectra 
(densities of states) of
the interactions $\Sigma^R(\tave,\xi_\kk)$ and the
Green's function $G^R(\tave,\xi_\kk)$. 
Note that on the right hand side, the term controlling
the dynamics is the difference in occupation functions
for $G$ and $\Sigma$, rather than $n_\kk(\tave)$ (which would have led to simple exponential
decay); in equilibrium, both occupation functions revert to the same Fermi-Dirac distribution function and the time derivative of the populations vanishes. This balance between the occupation functions
is the origin of much of the observed dynamics
and the mismatch of nonequilibrium dynamics from the equilibrium intuition.

This analysis, coupled with further theoretical work~\cite{lex_prx}, showed that
one or more of the population decay channels can saturate,
which resolves the experimental dilemma.
The most extreme example of this is impurity scattering, which
contributes to the line width but does not contribute to
the population decay rate. 
The underlying reason for this is
that any population decay must dissipate energy, and impurity
scattering (in the Born approximation) is elastic. Similar
considerations arise for electron-electron (el-el) scattering; the
role el-el scattering plays is proportional to how far an
electronic system is from a thermal distribution. Since the
electronic system can be at effective temperatures far above
the lattice (its thermal bath), electron-phonon dominated dynamics
continue even as the el-el dynamics shut off. This was experimentally
verified by Rameau \textit{et al.}, who showed that in a strongly correlated
system, the dynamics after the pump were dominated by electron-phonon
scattering \cite{rameau2016energy,konstantinova2018nonequilibrium}.  

The underlying reason for this is that the population
dynamics are in a sense determined by that of energy
transfer.  Elastic impurity scattering can at most
redistribute momentum, and thus once a momentum balance in the Brillouin
has been restored, it no longer plays a role.
Similarly, Coulomb scattering maintains the energy
within the electronic system, and thus can at most
lead to quasi-thermalization of the electronic subsystem.
However, electron-phonon scattering can take energy out
of the electronic populations, and thus determines
dynamics over a longer time scale. In fact, this can
lead to the electron-phonon coupling entirely dominating
the time dynamics even in strongly correlated systems,
as was shown by Rameau \textit{et al.}~\cite{rameau2016energy}
who demonstrated this for the high-$T_c$ cuprates.

\subsection{Changes in spectral shape}

The self-energy strongly depends on the populations,
and thus can acquire dynamics as well. This results in 
changes to the spectral line shape, which can be
resolved using trARPES~\cite{zhang2014ultrafast,rameau2014photoinduced,ishida2016quasi}. These observations were interpreted
as a decoupling of the electrons and phonons because
the kink in the quasiparticle bandstructure
softened. However, this is a place where equilibrium
intuition also fails. As was shown by Kemper \textit{et al.}~\cite{kemper2014effect}, the weakening of the kink can
simply be understood by observing that the sharp
structures in the electron-phonon self-energy
(c.f. Fig.~\ref{fig:sigmas}) become less sharp after a pump,
and this weakens the kink in the quasiparticle band structure.
Similarly, one can consider the effect of the pump to
produce a higher effective temperature for the electrons,
which would yield the same changes in the spectra.

\subsection{Excitons}

Recently, trARPES has been used to observe bound states
of electrons and holes, namely excitons~\cite{na2020probing}.
Here, while the overall
theoretical approach outlined above becomes more complex
in order to take into account two-particle bound
objects, the conceptual framework can be similar as that
developed by Freericks \textit{et al.}~\cite{freericks2009theoretical}.
\LK{However, unlike for the single-particle excitation spectra, the topic of observing
excitons with photoemission is still evolving; in addition to the simple questions,
complications arise due to the possibility of exciting virtual excitons or coherences~\cite{haug2009quantum}.
Several theoretical efforts have begun in this direction, showing that the 
resulting
spectra are a complex mixture of the valence and conduction
bands~\cite{rustagi2018photoemission,sangalli2018ab,perfetto2020time,christiansen2019theory,kemper2020observing,stefanucci2021carriers}.} 
Experimentally, in the two-dimensional
di-chalcogenide materials, where the exciton binding
energy is high, a direct excitation into the exciton
yields a clear signature in trARPES~\cite{buss2017ultrafast,tanimura2019dynamics,tanimura2020momentum,wallauer2020direct,madeo2020directly}.


\section{From NEGF to Quantum Boltzmann to N-temperature models}


Alternative formulations to the full non-equilibrium Green's
functions discussed above are the quantum Boltzmann equation
and the Boltzmann equation. The latter is essentially a rate equation,
where one counts the particles scattering into/out of a given state,
which has been used successfully to interpret population dynamics
in various settings~\cite{sobota2014ultrafast,na2020establishing};
however, along
the way several approximations are made that neglect some of
the potential effects discussed above. We will outline these
approximations, and indicate the effects of the approximation.
This brief outline is further detailed by
Kamenev~\cite{kamenev2011field}.

(i) The time axes are rotated to $\tave/\trel$. Similar to the above,
assume that the non-equilibrium
dynamics of the system are slow compared to the inverse energies of
the system; this lets us Fourier transform $\trel\rightarrow\omega$.
This approximation disregards time dynamics on fast time scales.

(ii) Perform a gradient expansion in $\frac{\partial}{\partial t}$ and $\frac{\partial}{\partial\omega}$, that is, keep terms
to linear order in a series expansion. This limits the ways that
the interactions can change. At this point, we have recovered the
\emph{quantum Boltzmann equation}. Here, the Dyson
equation's history kernel (that leads to non-Markovian
dynamics) has been removed via the gradient expansion,
and only single-time dynamics remains.

(iii) We assume that all the quasi-particles live on-shell.
In other words, the quasi-particle distribution is infinitely
sharply peaked around $\omega=\xi_\kk$. This neglects
non-trivial many-body effects---which can be quite significant---that play a role in the description of the quasi-particles.  At this
point, we find the \emph{Boltzmann equation}.

(iv) The Boltzmann equation, combined with considerations of
energy conservation and approximating the Fermi/Bose functions
for electrons and phonons, respectively, leads to the two-temperature 
model (as outlined in Allen~\cite{allen1987theory}). If one has more reservoirs involved in the dynamics, one can generalize this to N-temperature models, with one temperature for each reservoir.

One can see that the Boltzmann equation approach (leading ultimately to a rate-equation approach and N-temperature models) involves a significant number of approximations. It should never be viewed as a starting point for analysis, but rather should be thought of as an approximate description, when the approximations are warranted. For example, when we examine the exact many-body dynamics, we say that if the Green's function and the self-energy share the same distribution function, then the population dynamics vanish and the system stays in a steady state. This occurs even if the distribution functions are not Fermi-Dirac distributions. Hence, it brings in a complex interplay between the two different times. Rigorously N-temperature models can never be exact, but if the distribution functions are close to Ferm-Dirac distributions, they can be quite accurate. This is why we see N-temperature models used so successfully in the analysis of pump-probe experiments.

\section{Conclusions}

In this work, we provided a brief overview of how spectral response functions are modified in a nonequilibrium setting, especially in a pump-probe experiment. While this work only touched the surface, we provided a clear illustration of why one should not think of spectral responses as occurring at a definite time. Instead, they mix together the responses over a range of times, typically determined by the temporal spread of the probe function. While exact theory never allows a rigorous separation into time-dependent spectra, with well-defined times, there are situations where this becomes a quite accurate description. When the time variation along the average time is slow, or as we approach the long-time steady state, the system can be more and more accurately approximated by slowly varying spectra at ``definite'' average times. But, even in this case, the dynamical rules governing the decay of dynamics in the relative-time direction are often quite different from the decay dynamics along the average-time direction.  This helps explain some of the puzzling results that can arise when one tries to force equilibrium reasoning into a nonequilibrium setting. Oftentimes, it will not work perfectly. In many cases the issue is simply with misinterpreting the two-time behavior of the response functions. One should always look into that first to see if it will clear up misconceptions.

In this work, we also focused on describing how this behavior occurs in the context of tr-ARPES experiments. But the general principles have a much wider application. They govern the behavior of all two-time response functions.



\begin{acknowledgments}
J.~K.~F.~ was supported by the Department of Energy,
Office of Basic Energy Sciences, Division of Materials Sciences
and Engineering under Contract No. DE-FG02-08ER46542
(Georgetown). J.~K.~F.~ was also supported by the McDevitt bequest
at Georgetown.
A.~F.~K. acknowledges support from the National Science Foundation under Grant No. DMR-1752713. 
\end{acknowledgments}

\bibliography{refs}
\bibliographystyle{apsrev4-2}

\end{document}